\documentclass[%
 reprint,
 amsmath,amssymb,
 aps,
longbibliography,
]{revtex4}

\usepackage{graphicx}
\usepackage{float}

\usepackage{dcolumn}
\usepackage{bm}

\usepackage{tikz}
\usetikzlibrary{decorations.pathreplacing}

\begin{document}

\title{s- and p-superfluidity  of Fermi atoms in Bose-Fermi mixtures}


 \author{E.V. Gorbar$^{1,2}$, Y.O. Nikolaieva$^{1}$, I.V. Oleinikova$^{3}$, S.I. Vilchinskii$^{1}$,  A.I. Yakimenko$^{1}$}

\affiliation{$^1$  Department of Physics, Taras Shevchenko National University of Kyiv, 64/13, Volodymyrska Street, Kyiv 01601, Ukraine \\
$^2$ Bogolyubov Institute for Theoretical Physics, 14-b, Metrologichna, Kyiv 03143, Ukraine\\
$^3$ National University of Technologies and Design, 2,
Nemirovich-Danchenko Street, Kyiv 01011, Ukraine
}

\begin{abstract}
The $p$-wave superfluid is characterized by nontrivial topological characteristics essential for fault-tolerant
quantum state manipulation.
However, the practical realization of the $p$-wave state remains a challenging problem.
We study the $s$- and $p$-wave superfluidity in mixtures of fermionic and spinor bosonic
gases and derive a general set of the gap equations for these superfluid states. Numerically solving the gap equations for the $s$-wave state, we quantify the physical conditions for the realization of the pure $p$-wave state in a well-controlled environment of atomic physics in the absence of an admixture of the $s$-wave state.
\end{abstract}

\maketitle

\section{Introduction}

Chiral $p$-wave superfluids and superconductors have attracted much attention because of their rich physics and remarkable potential applications. Vortex excitations in
a two-dimensional version of such a condensate are anyonic and obey non-Abelian statistics.  Topologically degenerate quasiparticle states of chiral superconductors can be used for the creation of fault-tolerant topological quantum computers
\cite{RevModPhys.80.1083}. Cold atomic gases provide a promising physical platform for the realization of topologically non-trivial states.
The $p$-wave fermionic superfluidity in cold gases has become the topic of extensive theoretical research
\cite{PhysRevA.94.043632, PhysRevLett.117.245302,cooper2009stable,zhang2008p,julia2013engineering,han2009stabilization,bardyn2012majorana,2020PhRvA.101a3613T,sato2009non}. Although there are various $p$-wave Fermi superfluids such as superfluid phases of
$^3\mbox{He}$ \cite{Leggett} and heavy-fermion superconductors \cite{Sigrist,Stewart,Maeno}, the realization of $p$-wave superfluid state in a Fermi gas with
tunable pairing interaction would strongly advance the physics of unconventional Fermi superfluids both theoretically and experimentally.

The properties of fermionic pairing are determined by the agent, which mediates the effective attraction between fermions. In quantum gases, a Bose-Einstein condensate (BEC) can serve as such
a mediator in Bose-Fermi mixtures. Recent theoretical studies reveal rich physical properties of the fermi superfluidity in Bose-Fermi mixtures \cite{2020NatSR..1010822W,2018PhRvL.121z3001B,2018PhRvA..98e3630W}. 
 In Ref. \cite{PhysRevB.65.134519}, it was shown that {the} $p$-wave Cooper pairing of spin-polarized fermions (this makes
impossible the $s$-wave pairing) takes place in a binary fermion-boson mixture due to the exchange of density fluctuations of the bosonic
medium. The main problem is that the $s$-wave pairing (with zero orbital angular momentum of the pair) has typically a higher critical
temperature than the $p$-wave pairing and, therefore, is more favorable. As was demonstrated experimentally \cite{PhysRevLett.90.053201} in a single-component ultracold
Fermi gas of $^{40}$K atoms, {by using the Feshbach resonance, it is possible to essentially enhance the} $p$-wave collision cross-section to values
larger than the background $s$-wave cross-section between atoms in different spin states. However, the {Feshbach resonance} also significantly enhances three-body inelastic collisional loss \cite{Levinsen,Cooper,Pricoupenko}.

A spinor Bose-Einstein condensate (BEC) supports various types of excitations not available in single-component BECs, which offers considerable scope for the realization of unconventional superfluid states of quantum gases.
In our recent paper \cite{PhysRevA.98.043620}, we analyzed the $p$-wave polar phase in mixtures of ultracold Fermi and
$F=1$ Bose gases.
It was found that the spin-induced interactions in a 
mixture of spinor atomic BEC and fermionic atoms
naturally suppress the $s$-wave pairing. At the same time, the $p$-wave
pairing is driven both by density-induced and spin-induced interactions. Thus, in contrast to the $s$-wave state, the $p$-wave one emerges even for
repulsive interaction between fermions.
Although we demonstrated in Ref. \cite{PhysRevA.98.043620} that the polar $p$-wave phase can be realized, its competition with
the $s$-wave state was left as an open question. 

In the present work, we study the competition between the $s$- and $p$-wave superfluidity of the fermi atoms. We found numerical solutions to the gap equation in $s-$wave state and quantify the physical conditions for the realization of the pure $p$-wave state in the absence of an admixture of the $s$-wave state.

The rest of the paper is organized as follows. The model is introduced in Sec. II. In Sec. III, we derive the general gap equation and present the polar $p$-wave and $s$-wave gap equations. In Sec. IV,
we analyze the $s$-wave state and present our numerical results of the corresponding gap equation. The paper is
concluded in Sec. V.

\section{Model}

Our model of a mixture of electrically neutral $F=1$ bosonic and spin-$\frac12$ fermionic atoms is defined by the following second quantized Hamiltonian of dilute Fermi and spinor Bose
gases:
\begin{equation}\label{HamiltTotal}
\hat H=\int\left[\mathcal{\hat H}_B(\textbf{r})+\mathcal{\hat H}_F(\textbf{r})+\mathcal{\hat H}_{BF}(\textbf{r})\right]d\textbf{r}.
\end{equation}
The first term $\mathcal{\hat H}_{B}(\textbf{r})$ in the Hamilton density describes {interacting} $F=1$  spinor bosonic atoms with no external trapping potential
\cite{PhysRevLett.81.742,2012PhR...520..253K}
\begin{eqnarray}\label{HamiltB}\nonumber
\mathcal{\hat H}_{B}(\textbf{r})=\sum_m \hat{\Psi}_m^+(\textbf{r})\left(-\frac{\hbar^2}{2M_B}\nabla^2\right)\hat{\Psi}_m(\textbf{r}) + \frac12 c_0 : \hat n_B^2(\textbf{r}) :+\frac12 c_2  : \textbf{F} ^2:,
\end{eqnarray}
where $:\hat A:$ denotes normally ordered operator $\hat A$ and the field operators $\hat{\Psi}(\textbf{r})$ {obey} the standard commutation relations
$$
\left[\hat{\Psi}_m(\textbf{r}),\hat{\Psi}_{m'}^+(\textbf{r})\right]=\delta_{m, m'}\delta (\textbf{r}-\textbf{r}').
$$
Further,
$c_0=\frac{4\pi \hbar^2}{3 M_{B}}\left(a^{(0)}_{B}+2 a^{(2)}_{B}\right)$ and
$c_2=\frac{8\pi \hbar^2}{3 M_{B}}\left(a^{(2)}_{B} - a^{(0)}_{B}\right)$
are symmetric and spin-dependent interaction constants, respectively, where $a^{(J)}_{B}$ is the scattering length in the state with spin $J$. The operator
$\hat n_B(\textbf{r})=\sum_m \hat{\Psi}_m^+(\textbf{r}) \hat{\Psi}_m(\textbf{r})$, $m\in \{-1,0,+1\}$ denotes the number density for bosonic fields.
As to the ultracold gas of spin-1 bosonic atoms, we
consider {in this paper $^{23}$Na atoms in the polar spinor Bose--Einstein condensate (BEC) state with $c_2/c_0=6.25\cdot 10^{-2}$} and $a^{(0)}_{B}=50$, $a^{(2)}_{B}=55$ in units of the Bohr radius $a_B$.

The second term $\mathcal{\hat H}_{F}(\textbf{r})$ in the Hamilton density (\ref{HamiltTotal}) describes the gas of Fermi atoms without external trapping potential
\begin{equation}\label{HamiltF}
\mathcal{\hat H}_F(\textbf{r})=\sum_\nu \hat{ f }_\nu^+(\textbf{r})\left(-\frac{\hbar^2}{2M_F}\nabla^2-\mu \right)
\hat{ f }_\nu(\textbf{r})
 + \frac12 g_F : \hat n_F^2(\textbf{r}) :
\end{equation}
with the {field operators} $\hat{ f }$ obeying the standard anticommutation {relations}
$$
\left\{\hat{ f }_\nu(\textbf{r}),\hat{ f }_{\nu'}^+(\textbf{r})\right\}=\delta_{\nu, \nu'}\delta (\textbf{r}-\textbf{r}')
$$
and $\mu$ is the fermion chemical potential.
The operator $\hat n_F(\textbf{r})=\sum_\nu \hat{ f }_\nu^\dag(\textbf{r}) \hat{ f }_\nu(\textbf{r})$, $\nu\in \{\uparrow,\downarrow\}$ denotes the number density of Fermi atoms,
$g_F =  4\pi \hbar^2a_{F}/ M_{F}$ is the fermion-fermion coupling constant, where $a_{F}$ is the $s$-wave scattering length.

The last term $\mathcal{\hat H}_{BF}(\textbf{r})$ in the Hamilton density (\ref{HamiltTotal}) describes the interaction between the Bose and Fermi atoms
\begin{equation}\label{HBF}
\mathcal{\hat H}_{BF}(\textbf{r})=\frac{\alpha}{2}\, \hat n_B(\textbf{r}) \hat n_F(\textbf{r}) +\frac{\beta}{2}\, \textbf{S}\cdot \textbf{F},
\end{equation}
where {$\alpha = \frac{4\pi \hbar^2}{3 M_{BF}}\left(2a^{(1/2)}_{BF}+a^{(3/2)}_{BF}\right)$ and
$\beta = \frac{8\pi \hbar^2}{3 M_{BF}}\left(a^{(1/2)}_{BF}-a^{(3/2)}_{BF}\right)$ are} coupling constants of the density-density and spin-spin
interactions of Bose and Fermi atoms, respectively. Further, $a^{(J)}_{BF}$ is the scattering length in the state with spin $J$ and $M_{BF}=2M_B M_F/(M_B+M_F)$ is the reduced
mass for a boson with mass $M_B$ and a fermion with mass $M_F$.

The $j$-th component of the vector spin density operator for bosons $\textbf{F}$ equals
$$
\textbf{F}_j=\sum_{m, m'} \hat{\Psi}_m^+(\textbf{r}) \left\{\hat F_j\right\}_{m, m'}\hat{\Psi}_{m'}(\textbf{r}),
$$
where $\hat{F}_j$ {are the $3\times 3$ matrices of the vector representation of the rotation group}. The spin density operator for
fermions {has a similar form}
$$
\textbf{S}_j=\sum_{\nu, \nu'} \hat{ f }_\nu^\dag(\textbf{r}) \left\{\hat \sigma_j\right\}_{\nu, \nu'}\hat{ f }_{\nu'}(\textbf{r}),
$$
where $\hat\sigma_j$ are the Pauli-matrices for spin-$\frac12$ fermions.



To be specific, we fix in our study the fermionic and bosonic  densities $n_F=n_B = 10^{14}$ cm$^{-3}$ 
for $^{40}$K and $^{23}$Na atoms that gives the Fermi energy
$E_F = (3\pi^2 n_F)^{2/3}\hbar^2/(2 M_F) = \hbar^2 k_F^2/(2 M_F)$.


\section{Gap equations}


To derive the gap equations for both $s$- and $p$-wave states, it is convenient to obtain the
effective action for Fermi atoms. Using Hamiltonian (\ref{HamiltTotal}), {it is straightforward to show that the fermion-boson interactions
lead to} the following effective fermion's action:
\begin{equation}\label{eq62}
    S_{\text{eff}} = S_0 + S_{int},
\end{equation}
where
\begin{equation}
   S_0 = \int d\textbf{r}dt\Big[ \sum_{\nu} \hat{f}^+_{\nu}\left(i \hbar \frac{\partial}{\partial t} +\frac{\hbar^2}{2m}\nabla^2
   + \mu \right)\hat{f}^+_\nu - \frac{1}{2}g_F : \hat n_F^2(\textbf{r}) :\Big].
\end{equation}
is the free action of Fermi atoms and the interaction term is given by
\begin{equation}
   S_{int} = -\sum_{\textbf{k}\neq0}\frac{\epsilon_{\textbf{k}}n_B}{2}\int d\omega\Big\{\frac{\beta^2 }{E_{\textbf{k},1}^2
   - \hbar^2\omega^2} [\sigma_+(k)\sigma_-(-k)+\sigma_-(k)\sigma_+(-k)]
   +\frac{\alpha^2  }{2 (E_{\textbf{k},0}^2 - \hbar^2\omega^2)}n_F(k)n_F(-k)\Big\}.\\
\end{equation}
Here $\epsilon_{\textbf{k}} = \hbar^2\textbf{k}^2/(2 M_B)$,
$k=(\omega,\mathbf{k})$, $r=(t,\mathbf{r})$,
\begin{equation}\label{sigmaPM}
   \sigma_\pm=\frac{1}{2\sqrt{\Omega}}\int e^{i \textbf{kr}}\hat f^+(\textbf{r})(\hat{\sigma}_x\pm i \hat{\sigma}_y)\hat
   f (\textbf{r})d \textbf{r},\\
   \end{equation}
   \begin{equation}\label{BdG}
     E_{\bm{k},0} = \sqrt{\epsilon_{\bm{k}} (\epsilon_{\bm{k}} + 2c_0 n_B)},\quad
   E_{\bm{k},1} = \sqrt{\epsilon_{\bm{k}} (\epsilon_{\bm{k}} + 2c_2 n_B)}, 
   \end{equation}
and $\Omega$ is the system's volume.

The effective action \eqref{eq62} {implies the following Dyson's equations}:
\begin{equation}\label{Geq3}
    (i\hbar\omega_n-\xi_{\textbf{k}})G(\textbf{k},\omega_n)+\Delta(\textbf{k},\omega_n)F^+(\textbf{k},\omega_n) = \hbar,
\end{equation}
\begin{equation}\label{F+eq3}
    (i\hbar\omega_n+\xi_{\textbf{k}})F^+(\textbf{k},\omega_n)+\Delta^+(\textbf{k},\omega_n)G(\textbf{k},\omega_n)=0,
\end{equation}
for the fermion Green function $G(\textbf{k},\omega_n)$ and
the anomalous Gor`kov function $F(\textbf{k},\omega_n)$ in the
Matsubara formalism,
where $\xi_\mathbf{k} = \hbar^2\textbf{k}^2/(2M_F)-\mu$. Here we introduced the order parameter of general type
\begin{equation}\label{GAP}
\Delta
    =-\frac{k_\textrm{B} T}{\hbar}\sum_{\nu,\textbf{k}'\neq 0}\Big[ \frac{g_F}{\Omega}F(\textbf{k}-\textbf{k}',\omega_\nu) -
    \tilde{g}_n (\textbf{k}',\omega_\nu)  F(\textbf{k}-\textbf{k}',\omega_n - \omega_\nu)
    -\tilde{g}_s (\textbf{k}',\omega_\nu) \sum_{s = +,-}\hat{\sigma}_s F(\textbf{k}-\textbf{k}',\omega_n - \omega_\nu)\hat{\sigma}_s
    \Big],
\end{equation}
where
\begin{align*}
    \tilde{g}_n(\textbf{k},\omega_\nu) =  \frac{c_n}{ E_{\textbf{k},0}^2 + \hbar^2\omega_\nu^2} , &\qquad
    c_n = \frac{\alpha^2 n \epsilon_{\textbf{k}}}{2\Omega},\\
    \tilde{g}_s (\textbf{k},\omega_\nu)=  \frac{c_s}{ E_{\textbf{k},1}^2 + \hbar^2\omega_\nu^2} ,&\qquad
    c_s = \frac{\beta^2 n \epsilon_{\textbf{k}}}{\Omega}.
\end{align*}

Since we wish to describe both $s$- and $p$- wave states, we
consider the following ansatz for the energy gap:
\begin{equation}\label{gapsi}
    \Delta_{\alpha \beta} (\textbf{k}) = \Delta_p (i\bm{\sigma} \sigma_y)_{\alpha \beta} \textbf{d}(\textbf{k})
    + \Delta_s (i\sigma_y)_{\alpha\beta}.
\end{equation}
The vector $\textbf{d}$ determines the form and symmetries of the p-wave superfluid state. Further, $\alpha,\beta$ are the spin indices, $\bm{\sigma}$ are the Pauli matrices, and $\Delta_s$ and $\Delta_p$ are scalar functions of $\omega_n$ and $|\mathbf{k}|$. Using this ansatz,
Eqs. \eqref{Geq3} and \eqref{F+eq3} give
\begin{equation}\label{solGPS}
    G_{\alpha\beta}(\textbf{k},\omega_n) = - (i\hbar\omega_n +\xi_{\textbf{k}})\cdot \frac{(\hbar^2\omega_n^2+\xi_{\textbf{k}}^2
    +\Delta^2)\delta_{\alpha\beta} - \Delta^2\textbf{M}\bm{\sigma}_{\alpha\beta}}{(\hbar^2\omega_n^2 + E_+^2)(\hbar^2\omega_n^2 + E_-^2)},
\end{equation}
\begin{eqnarray}\label{solFPS}
    F_{\alpha\beta}(\textbf{k},\omega_n) &=& \hbar\Delta_p (\bm{\sigma}i\sigma_y)_{\alpha\beta}\cdot \frac{(\hbar^2\omega_n^2
    +\xi_{\textbf{k}}^2+\Delta^2)\textbf{d} - i\Delta^2[\textbf{M}\cdot\textbf{d}]}{(\hbar^2\omega_n^2 + E_+^2)(\hbar^2\omega_n^2 + E_-^2)}\nonumber\\
    &&+\hbar\Delta_s\frac{(\hbar^2\omega_n^2+\xi_{\textbf{k}}^2+\Delta^2-2\Delta_p^2(Re[\textbf{d}]\cdot\textbf{d}))i\sigma_{y,\alpha\beta} }
    {(\hbar^2\omega_n^2 + E_+^2)(\hbar^2\omega_n^2 + E_-^2)}+\hbar\Delta_s\frac{ \Delta^2(\bm{\sigma}i\sigma_y)_{\alpha\beta}\textbf{M}}
    {(\hbar^2\omega_n^2 + E_+^2)(\hbar^2\omega_n^2 + E_-^2)},
\end{eqnarray}
where
\begin{gather*}
    \Delta^2 = \Delta_p^2 |\textbf{d}|^2 +\Delta_s^2, \\
    \textbf{m}(\textbf{k}) = i [\textbf{d}(\textbf{k})\times \textbf{d}^*(\textbf{k})],\\
    \textbf{M} = \frac{\Delta_p^2\textbf{m} + 2 \Delta_p\Delta_s Re[\textbf{d}]}{\Delta^2},\\
    E_{\pm} =\sqrt{\xi_{\textbf{k}}^2 + \Delta^2 ( 1 \pm | M |) }.
\end{gather*}

For the polar $p$-wave state with $\mathbf{d}=(0,0,k_z)$, Eqs.(\ref{GAP}) and (\ref{solFPS}) lead to the following $p$-wave gap equation:
\begin{equation}\label{p-gap-equation}
    k_z\Delta_p(\textbf{k},\omega_k) = k_\textrm{B} T\,\sum_{n,\textbf{k}'\neq0}( \tilde{g}_n(\textbf{k}',\omega_n)
    + \tilde{g}_s(\textbf{k}',\omega_n))\,\, \frac{(k_z-k'_z)\Delta_p(\textbf{k}-\textbf{k}',\omega_k-\omega_n)}{\hbar^2(\omega_k-\omega_n)^2+\xi^2_{\textbf{k}-\textbf{k}'}+(k_z-k'_z)^2\Delta^2_p}.
\end{equation}
The same equations for the s-wave gap $\Delta_s$ give the following gap equation:
\begin{equation}\label{s-gap-equation}
    \Delta_s(\textbf{k},\omega_k) = k_\textrm{B} T\, \sum_{n,\textbf{k}'\neq0}\left(-\frac{g_F}{\Omega} + \tilde{g}_n(\textbf{k}',\omega_n) - \tilde{g}_s(\textbf{k}',\omega_n)\right)\,\,\frac{\Delta_s(\textbf{k}
    -\textbf{k}',\omega_k-\omega_n)}{\hbar^2(\omega_k-\omega_n)^2+\xi^2_{\textbf{k}-\textbf{k}'}+\Delta^2_s}.
\end{equation}


\subsection{p-wave gap equation}


Although we concentrate in this paper on the $s$-wave state, it is useful to present here the details of the derivation of the $p$-wave gap equation obtained in Ref. \cite{PhysRevA.98.043620}.
Setting $\omega_k=\omega_0=\pi k_\textrm{B}T/\hbar$ in Eq.(\ref{p-gap-equation}) and neglecting the dependence of the gap on the Matsubara frequency, we obtain the following equation for the polar phase gap
$\Delta(\bm{k})=k_z\Delta_p(\mathbf{k},\omega_0)$:
\begin{equation}\label{solz}
    \Delta(\bm{k}) = k_\textrm{B} T\cdot \sum_{n,\textbf{k}'\neq0}( \tilde{g}_n(\textbf{k}',\omega_n-\omega_0)
    + \tilde{g}_s(\textbf{k}',\omega_n-\omega_0))\cdot\frac{\Delta(\textbf{k}-\textbf{k}')}{\hbar^2\omega_n^2 + E^2(\textbf{k}-\textbf{k}^{\prime})}.
\end{equation}
By using the definition of $\tilde{g}_n$ and $\tilde{g}_s$ below Eq.(\ref{GAP}), we conclude that the summation over $n$ boils down to the calculation of the following sum with certain $a$ and $b$:
\begin{eqnarray}\nonumber
 S = k_\textrm{B}T\sum_n \frac{1}{\pi^2 k_\textrm{B}^2T^2 (2n + 1)^2 + a^2}\cdot \frac{1}{\pi^2 k_\textrm{B}^2T^2 (2n + 1 - k)^2 + b^2} \\ = k_\textrm{B}T\sum_{n,s1,s2}\frac{s_1 s_2}{4 a b} \frac{1}{i \pi k_\textrm{B}T (2 n + 1) - s_1 a}\cdot \frac{1}{i \pi k_\textrm{B} T(2n + 1) - i \pi k_\textrm{B} T k - s_2 b},
\end{eqnarray}
which can be explicitly calculated, where we introduced auxiliary variables $s_1,s_2$ with values $\pm 1$. Indeed, we have
\begin{equation}
k_\textrm{B}T \sum \frac{1}{(i \hbar\omega_n - \xi_1)(i\hbar\omega_n - \xi_2)} = \frac{n_F(\xi_1) - n_F(\xi_2)}{\xi_1 - \xi_2},
\end{equation}
where $
\xi_1 = s_1 a$, and $\xi_2 = s_2 b + i \pi k_\textrm{B}T k$. Thus, we obtain
\begin{equation}
S = \sum_{s_1,s_2} \frac{s_1 s_2}{4 a b} \frac{n_F(s_1 a) + n_B(s_2 b)}{s_1 a - s_2 b - i \hbar\omega},
\end{equation}
where we used
\begin{equation}
n_F(i \pi  k_\textrm{B}Tk + x) = -n_B(x), \qquad k = 2 q + 1.
\end{equation}
Replacing $s_2 \to -s_2$, we have
\begin{equation}
S = \sum_{s1,s2} -\frac{s_1 s_2}{4 a b} \frac{-n_B(-s_2 b) - n_F(s_1 a)}{i \hbar\omega - s_1 a - s_2 b} = -\sum_{s1,s2} \frac{s_1 s_2}{4 a b} \frac{1 + n_B(s_2 b) - n_F(s_1 a)}{i \hbar\omega - s_1 a - s_2 b}.
\end{equation}

Using 
\begin{equation}
n_F(x) = \frac{1 - \tanh \frac{x}{2 k_\textrm{B}T}}{2}, \qquad n_B(x) = \frac{\coth\frac{x}{2k_\textrm{B} T} - 1}{2},
\end{equation}
we can rewrite
\begin{equation}
S = \sum_{s_1,s_2} \frac{-1}{8 a b} \frac{s_1\coth \frac{b}{2 k_\textrm{B}T} + s_2 \tanh \frac{a}{2 k_\textrm{B}T}}{i \hbar\omega - s_1 a - s_2 b}.
\end{equation}
Let us calculate first
\begin{eqnarray}\nonumber
\sum_{s1,s2} \frac{-1}{8 a b} \frac{(i \hbar\omega + s_1 a - s_2 b)s_1 \coth \frac{b}{2 k_\textrm{B}T}}{(i \hbar\omega - s_2 b)^2 - a^2} = \sum_{s_2}\frac{-1}{4 b} \frac{\coth \frac{b}{2 k_\textrm{B}T}}{-a^2 - \hbar^2\omega^2 + b^2 - 2 i \hbar\omega s_2 b} \\ = \frac{\coth \frac{b}{2 k_\textrm{B}T}}{2 b} \frac{\hbar^2\omega^2 + a^2 - b^2}{(\hbar^2\omega^2 + a^2 - b^2)^2 + 4 \hbar^2\omega^2 b^2}.
\end{eqnarray}
Similarly we have
\begin{equation}
S =\frac{\coth \frac{b}{2 k_\textrm{B}T}}{2 b} \frac{\hbar^2\omega^2 + a^2 - b^2}{(\hbar^2\omega^2 + a^2 - b^2)^2 + 4 \hbar^2\omega^2 b^2} + \frac{\tanh \frac{a}{2 k_\textrm{B}T}}{2 a} \frac{\hbar^2\omega^2 + b^2 - a^2}{(\hbar^2\omega^2 + b^2 - a^2)^2 + 4 \hbar^2\omega^2 a^2}.
\end{equation}
According to Eq.(\ref{solz}),
\begin{equation}
b = E_{k - k'}, \qquad a = E_{\mathbf{k},i},\qquad i=\{0,1\},
\end{equation}
therefore, we have
\begin{equation}
S = \frac{Z_i}{2 E_{\mathbf{k},i}} \tanh \frac{E_{\mathbf{k},i}}{2 k_\textrm{B}T} + \frac{W_i}{2 E_{k - k'}} \coth \frac{E_{k - k'}}{2k_\textrm{B} T},
\end{equation}
where
\begin{align}
Z_i = \frac{\hbar^2\omega^2 + E_{k - k'}^2 - E_{\mathbf{k},i}^2}{(E_{k - k'}^2 - E_{\mathbf{k},i}^2)^2 + 2 \hbar^2\omega^2\left(E_{\mathbf{k},i}^2 + E_{k - k'}^2 + \frac{\hbar^2\omega^2}{2}\right)},
\label{Zi}\\
W_i = \frac{\hbar^2\omega^2 + E_{\mathbf{k},i}^2 - E_{k - k'}^2}{(E_{k - k'}^2 - E_{\mathbf{k},i}^2)^2 + 2 \hbar^2\omega^2\left(E_{\mathbf{k},i}^2 + E_{k - k'}^2 + \frac{\hbar^2\omega^2}{2}\right)}.
\label{Wi}
\end{align}
Finally, we obtain
\begin{equation}
\label{dz_correct}
\Delta(\bm{k}) = \sum_{\bm{k'}\neq0} \Big\{c_n\left(\frac{W_0}{2E_{k - k'}}\coth{\frac{E_{k - k'}}{2k_\textrm{B}T}}
+\frac{Z_0}{2E_{\bm{k},0}}\tanh{\frac{E_{\bm{k},0}}{2k_\textrm{B}T}}\right)
+c_s\left(\frac{W_1}{2E_{k - k'}}\coth{\frac{E_{k - k'}}{2k_\textrm{B}T}}+\frac{Z_1}{2E_{\bm{k},1}}\tanh{\frac{E_{\bm{k},1}}{2k_\textrm{B}T}}\right)\Big\}
\cdot \Delta(\bm{k}-\bm{k'}). 
\end{equation}

In what follows, we study the $s$-wave phase in detail because the $p$-wave polar state was previously investigated in our recent work \cite{PhysRevA.98.043620} including numerical solution of the corresponding gap equation.
\subsection{s-wave gap equation}

Similarly setting $\omega_k=\omega_0=\pi k_\textrm{B}T/\hbar$ in Eq.(\ref{s-gap-equation}) and neglecting the dependence of the gap on the Matsubara frequency, we obtain the following equation for the $s$-wave gap:
\begin{multline}
\label{s_wave_correct}
\Delta(\bm{k}) = \sum_{\bm{k'}\neq0} \Big\{-\frac{g_F}{\Omega} \frac{\tanh \frac{E_{k - k'}}{2 k_\textrm{B}T}}{2 E_{k - k'}} + c_n\left(\frac{W_0}{2E_{k - k'}}\coth{\frac{E_{k - k'}}{2k_\textrm{B}T}}
+\frac{Z_0}{2E_{\bm{k},0}}\tanh{\frac{E_{\bm{k},0}}{2k_\textrm{B}T}}\right)\\
- c_s\left(\frac{W_1}{2E_{k - k'}}\coth{\frac{E_{k - k'}}{2k_\textrm{B}T}}+\frac{Z_1}{2E_{\bm{k},1}}\tanh{\frac{E_{\bm{k},1}}{2k_\textrm{B}T}}\right)\Big\}
\cdot \Delta(\bm{k}-\bm{k'}),
\end{multline}
where we used also 
the Poisson summation formula 
with the subsequent extension of the 
contour to infinity)
$$
\sum^{+\infty}_{n=-\infty}\frac{1}{k_\textrm{B}^2T^2\pi^2(2n+1)^2+a^2}=\frac{\tanh(\frac{a}{2k_\textrm{B}T})}{2k_\textrm{B}Ta}
$$
and $W_i$, $Z_i$ are defined in Eqs.(\ref{Zi}) and (\ref{Wi}).
Now let us find solution to the gap equation of the $s$-wave state.

\section{Results for the s-wave state}\label{sec:s-wave}

Before we find the solution to the gap equation for the $s$-wave state Eq.(\ref{s_wave_correct}), it is useful to check the correctness of the obtained $s$-wave gap equation by considering the BCS limit.



\subsection{BCS limit}

Equation (\ref{s-gap-equation}) in the limiting case of vanishing bose-fermionic coupling ($\alpha\to 0$, $\beta\to 0$) yields the standard
BCS gap equation for attractive interactions in the fermionic sector ($g_F<0$)
\begin{equation}
\label{BCSeq}
\Delta(\mathbf{k})=\sum\limits_{\mathbf{k'} \neq 0}  -\frac{g_F}{\Omega}\frac{\tanh\frac{E_{\mathbf{k}-\mathbf{k'}}}{2 k_\textrm{B}T }} {2E_{\mathbf{k}-\mathbf{k'}}}\,\, \Delta(\mathbf{k}-\mathbf{k'}).
\end{equation}

In the zero temperature limit $T\rightarrow 0$, we obtain the well-known BCS gap equation at zero temperature
\begin{equation}
    \Delta = -\frac{ \Delta g_F}{2(2\pi)^3}\int\frac{d\textbf{k}}{\sqrt{\xi^2+\Delta^2}}= - \frac{ \Delta g_F M_F p_F}{2 \pi^2 \hbar^3}
    \ln \frac{\hbar\omega_D}{\Delta},
\end{equation}
where $\hbar\omega_D$ is the maximum exchange energy. For $g_F > 0 $, the only possible solution is trivial solution $\Delta = 0$. However,
if $g_F < 0$, we obtain the well known BCS-type result
\begin{equation*}
    \Delta = \hbar\omega_D e^{-\frac{1}{\zeta}},
\end{equation*}
{where $\zeta = | g_F | M_F p_0/(2 \pi^2 \hbar^3).$}

For a repulsive fermion-fermion interaction ($g_F\ge 0$), the 
electron-electron pairing is possible only due to the effective 
interaction induced by Bogoluibov excitations in the bosonic sector. 
For $g_F > 0$, the $s$-wave pairing could be realized only if $\tilde{g}_n$ is sufficiently large. The $s$-wave gap equation for
the $s$-wave channel implies the following inequalities for the interaction kernel:
\begin{multline}
    -\frac{g_F}{\Omega} + \tilde{g}_n(\textbf{k}',\omega_k-\omega_n) - \tilde{g}_s(\textbf{k}',\omega_k-\omega_n)
    \leq -\frac{g_F}{\Omega}
    + \tilde{g}_n(\textbf{k}',\omega_k-\omega_n)
    \leq -\frac{g_F}{\Omega} + \tilde{g}_n(0,\textbf{0})=-\frac{g_F}{\Omega}
    +\frac{\alpha^2}{4\Omega n c_0}.
\end{multline}
It is obvious that the $s$-wave channel is suppressed in systems with $g_F > 0$.

It is remarkable that the density-density bose-fermi interaction (the term proportional to $\alpha^2$ in Eq. \eqref{s-gap-equation})
produces effective attractive interactions in the $s$-wave channel. At the same time the spin-spin bose-fermi interactions (the term
proportional to $\beta^2$ in Eq. \eqref{s-gap-equation}) lead to an additional \textit{repulsive} interaction for the $s$-wave channel.
However, in sharp contrast to the $s$-wave superfluidity, for the $p$-wave superfluidity, both spin-spin and
density-density bose-fermi interactions lead to an effective attraction that results in the realization of the $p$-wave
superfluid.

\begin{figure}[htb]
\begin{center}
\includegraphics[width=15.cm]{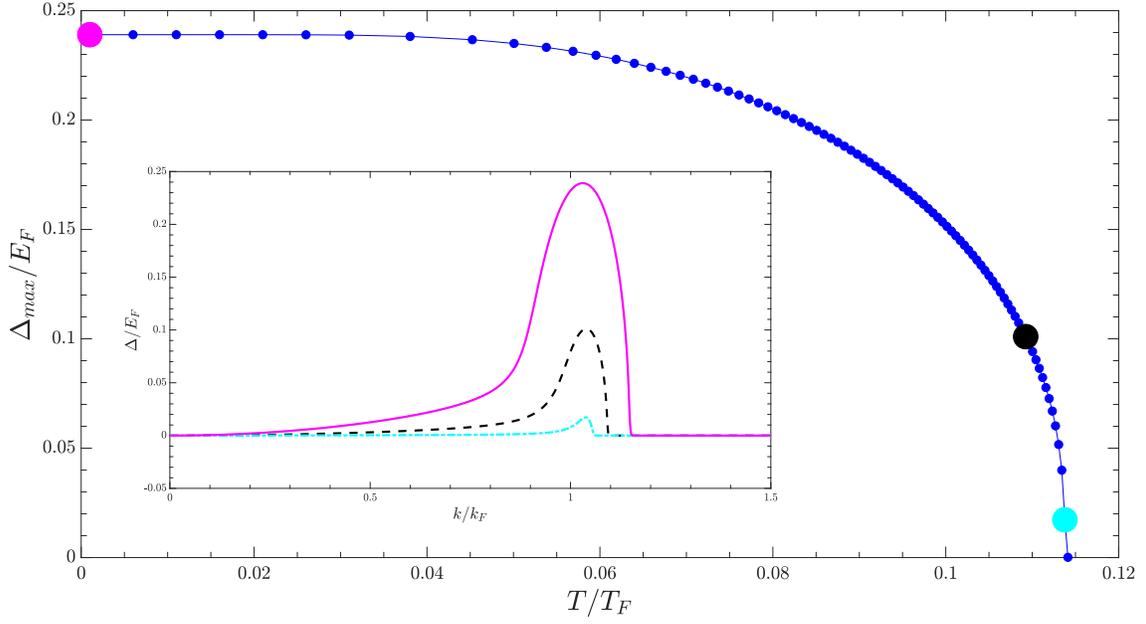}
\end{center}
\caption{The maximum value of the energy gap $\Delta_\textrm{max}$ of the $s$-wave state  (in units of the Fermi energy $E_F$) as a function of temperature $T$ (in units of the Fermi temperature $T_F=E_F/k_\textrm{B}$) found numerically in the BCS limit $\alpha=\beta=0$. The inset represents the energy gap $\Delta$ (in units of the Fermi energy $E_F$) as a function of momentum $k$ (in units of the Fermi momentum $k_F=(3\pi^2n_F)^{1/3}$) for temperature values indicated by colored circles on the $\Delta_\textrm{max}$ curve.}
\label{fig:BCS}
\end{figure}
\begin{figure}[htb]
\begin{center}
\includegraphics[width=15.cm]{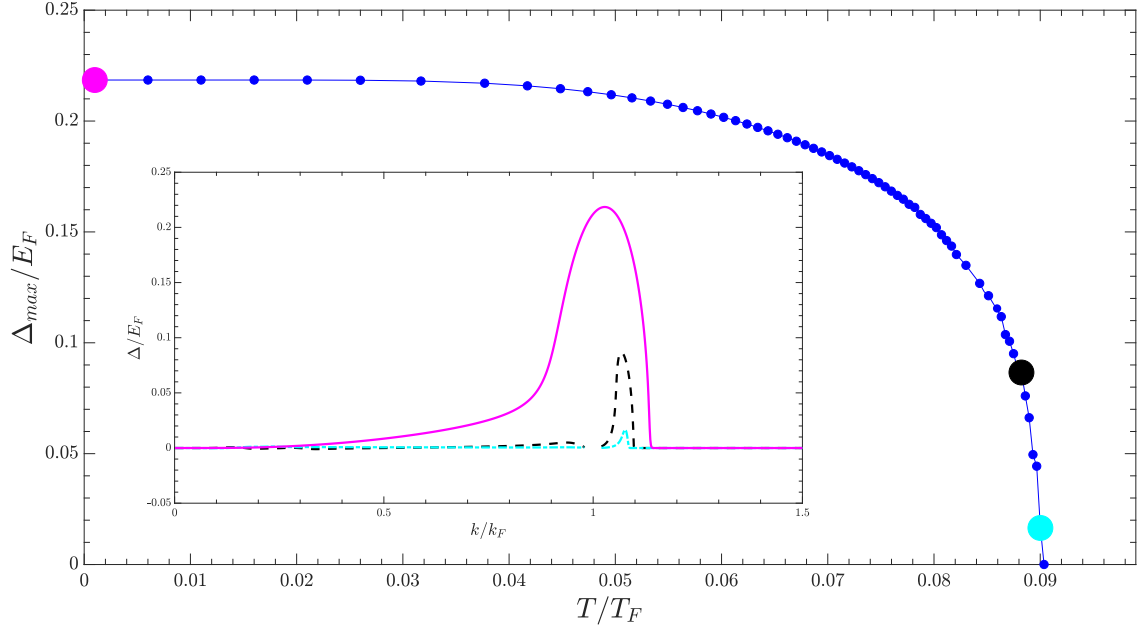}
\end{center}
\caption{The same as in Fig. \ref{fig:BCS} for $\alpha=0.3 c_0$, $\beta=0.3 c_0$.}
\label{fig:general}
\end{figure}

\subsection{Numerical solution}

We solve the gap equation for the $s$-wave state (\ref{s_wave_correct}) by using the following approximation for momentum-dependent functions:
\begin{equation}
\label{scf}
f(\mathbf{k}-\mathbf{k'})=f(k)\mathbf{\theta}(k-k')+f(k')\mathbf{\theta}(k'-k),
\end{equation}
where $\mathbf{\theta}$ is the Heaviside step function and
\begin{equation}\label{TKE}
   E_k=\sqrt{ \xi ^2_k+\Delta(k)^2}.
\end{equation}


To solve the gap equation (\ref{s_wave_correct}) numerically, we use the following stabilized relaxation procedure:
\begin{equation}
\label{NUMmaineq}
\Delta^{(n+1)}(\mathbf{k})=\int \mathcal{K}^{(n)}\times \Delta^{(n)}(\mathbf{k}-\mathbf{k'})d\mathbf{k},
\end{equation}
\begin{multline}
\mathcal{K}^{(n)}= -\frac{g_F}{\Omega}\frac{\tanh\frac{E_{\mathbf{k}-\mathbf{k'}}}{2 k_\textrm{B}T}} {2E_{\mathbf{k}-\mathbf{k'}}}
+c_n \left(  \frac{W_0}{2E_{\mathbf{k}-\mathbf{k'}}} \coth \frac{E_{\mathbf{k}-\mathbf{k'}}}{2 k_\textrm{B}T}+
\frac{Z_0}{2E_{\mathbf{k},0}} \tanh \frac{E_{\mathbf{k},0}}{2 k_\textrm{B}T}\right)
\\-c_s \left(  \frac{W_1}{2E_{\mathbf{k}-\mathbf{k'}}} \coth \frac{E_{\mathbf{k}-\mathbf{k'}}}{2 k_\textrm{B}T}+
\frac{Z_1}{2E_{\mathbf{k},1}} \tanh \frac{E_{\mathbf{k},1}}{2 k_\textrm{B}T}\right).
\end{multline}
The subsequent iteration is calculated by using the above equation multiplying $\Delta^{(n+1)}(\mathbf{k})$ by the ratio $(\mathcal{N}_n/\mathcal{N}_{n+1})^{1/2}$, where $\mathcal{N}_n=\int  \Delta^{(n)}(\mathbf{k}-\mathbf{k'})d\mathbf{k}$. 
Note that there are no solutions of the $s$-wave gap equation for repulsive fermi-fermi interactions ($g_F>0$) for considered here parameters $\alpha$ and $\beta$. Furthermore, for attractive fermi-fermi interactions ($g_F<0$) there is the threshold value of the interaction constant $g_F$ which allows the existence of $s$-wave superfluid state.

Our numerical results are presented in Figs.\ref{fig:BCS} and \ref{fig:general}.
The inset in Fig. \ref{fig:BCS} illustrates typical gap profiles for the case $\alpha=\beta=0$ as a function of momentum at different temperature. As one can expect, the maximum is located near the Fermi momentum $k=k_F$. The gap amplitude decreases with temperature and vanishes at the critical point.
Figure \ref{fig:general} shows typical solutions of the gap equation with nonzero density-density and spin-spin interactions for $\alpha=\beta=0.3c_0$. Notice that the numerical method used in the present work does not allow us to determine the gap values near the $k=k_F$ as is illustrated in the inset of Fig. \ref{fig:general}. Nevertheless, it is still possible to find the amplitude value as a function of temperature and determine the critical temperature when the pairing gap vanishes.
Note that spin-spin and density-density interactions decrease of the energy gap of the $s$-wave state.

\section{Conclusions}

The $s$- and $p$-wave superfluidity is investigated in a mixture of fermionic $^{40}$K and $F=1$ spinor bosonic $^{23}$Na
gases. As is known, in common superfluids, the $s$-wave phase dominates because it has a higher critical temperature. In the present work, we focused on the problem of the existence of the $s$-wave superfluid phase of fermions in Bose-Fermi mixtures and determined the physical conditions when the $p$-wave state is more favorable.
 The gap equations for the $s$-wave and the polar phase of $p$-wave superfluid fermions are derived. 
The gap equation for the $s$-wave state is solved numerically for different values of temperature. As one can expect, the maximum value of the gap is located near the Fermi momentum $k=k_F$ and the energy gap of the $s$-wave state is suppressed by spin-spin interactions of bosonic and fermionic atoms.
The physical conditions necessary for the realization of the $s$-wave and the polar phase in a well-controlled environment of atomic physics are determined.  Our analysis shows that a nontrivial solution to the $s$-wave gap equation is absent for repulsive fermi-fermi interactions ($g_F>0$) in the case of physically relevant parameters. As to attractive fermi-fermi interactions ($g_F<0$), the $s$-wave superfluid state exists only for sufficiently strong fermion-fermion interaction strength $g_F$. Thus, the $s$-wave state is suppressed compared to the $p$-wave state in Bose-Fermi mixtures. We hope that the obtained results can be useful for the practical realization of the $p$-wave state in ultracold atomic gases.



\section*{ACKNOWLEDGMENTS}
The authors are grateful to O.V. Bugaiko for the participation in the initial stage and substantial contributions to this work. The work of E.V.G. was partially supported by the Program of Fundamental Research of the Physics and Astronomy Division of the
National Academy of Sciences of Ukraine. Y.O.N. and A.I.Y. acknowledge partial support from the National Research Foundation of
Ukraine through Grant No. 2020.02/0032.


\bibliography{p_wave}
    
\end{document}